\documentclass[prb,preprint,showpacs,superscriptaddress,floatfix]{revtex4}
\usepackage{graphicx,color}
\usepackage[english]{babel}
\usepackage{dcolumn}

\begin{document}

\title{
  Cross-sectional imaging of\\
 sharp Si interlayers  embedded in gallium arsenide
}
\author{Xiangmei Duan}
\affiliation{INFM-DEMOCRITOS National Simulation Center, 
via Beirut 2-4, 34014 Trieste, Italy}
\affiliation{Institute of Solid State Physics, Chinese Academy of Sciences, 
  Hefei, China}
\author{Stefano Baroni}
\affiliation{INFM-DEMOCRITOS National Simulation Center, 
via Beirut 2-4, 34014 Trieste, Italy}
\affiliation{SISSA -- Scuola Internazionale Superiore di Studi
  Avanzati, via Beirut 2-4, 34014 Trieste, Italy}
\author{Silvio Modesti}
\affiliation{INFM-TASC National Laboratory, Area Science Park, 34012 
Trieste, Italy}
\affiliation{Dipartimento di Fisica, Universit\`a di Trieste, 
via Valerio 2, 34127 Trieste,
  Italy}
\author{Maria Peressi}
\email{peressi@ts.infn.it}
\affiliation{INFM-DEMOCRITOS National Simulation Center, 
via Beirut 2-4, 34014 Trieste, Italy}
\affiliation{Dipartimento di Fisica Teorica, Universit\`a di Trieste,
Strada Costiera 11, 34014  Trieste, Italy}
\date{\today}

\begin{abstract}
We investigate the electronic properties of the 
(110) cross-sectional surface of Si-doped GaAs 
using first-principles techniques.
We focus on doping configurations with an equal concentration of 
Si impurities in cationic and anionic sites,
such as occurring
in a self-compensating doping regime.
In particular we study a bilayer of Si atoms 
uniformly distributed over two consecutive (001) atomic layers.
The simulated cross-sectional scanning tunneling microscopy images 
show a bright 
signal at negative bias, which is strongly attenuated when the bias is
reversed.
This scenario is consistent with experimental results
which had been attributed to hitherto unidentified Si complexes.
\end{abstract}

\maketitle

Si doped GaAs has an
outstanding importance in the technology of electronic devices. 
Remarkably, Si in GaAs has an amphoteric behavior---being able to
substitute both arsenic and gallium atoms as an acceptor or a donor
respectively---resulting in a compensation mechanism which 
depends on the growth conditions and has 
been long debated.\cite{ModestiPRL04,Ebert-review}
Recently, GaAs samples with Si $\delta$-doped layers embedded in the
(001) direction 
have started to be investigated by cross-sectional scanning-tunneling 
microscopy and spectroscopy (XSTMS)
 on the (110) easy-cleavage surface.\cite{ModestiPRL04,Feenstra} 
It was found that in samples grown at 600$^\circ$C
Si self-compensation occurs above a surface Si concentration
of about 0.06 monolayer (ML) and 
involves nucleation and growth of
electrically neutral Si precipitates at the expense of the conventional
donor Si phase.\cite{ModestiPRL04}
In the XSTM images of such samples, beside the common and  well identified
defects such as silicon donors ($\rm Si_{\rm Ga}$) and gallium vacancies
($\rm V_{\rm Ga}$), new features appear,
bright at negative bias and strongly attenuated when the bias is
reversed.
The microscopic picture of these features is still unidentified, although
it is clear that they cannot be attributed  to
$\rm Si_{\rm Ga}$ donors only---whose STM images are well characterized
and remarkably different---nor to substitutional-vacancy complexes such as  
$\rm Si_{\rm Ga}$-$\rm V_{\rm Ga}$ or $\rm Si_{\rm Ga}$-$\rm V_{\rm
  As}$, as previously proposed.\cite{Ebert-review} 
This kind of contrast is also peculiar of the XSTM image 
of a bilayer of Si in GaAs, as shown in Fig. 1. 

A  few  theoretical studies of Si-doped or otherwise defected
GaAs (110) surfaces
exist,\cite{WangPRB93,LengelPRL94,CapazPRL95,ZhangPRL96,KimChelikowsky,EbertAPL01} 
but
to the best of our knowledge no attempts have been done so far to
model extended self-compensating donor-acceptor configurations. The
purpose of the present paper is to start filling this gap 
by providing the first theoretical study of XSTM images
resulting from the cleavage
of (001) Si interlayers embedded in GaAs. 
When Si substitutes both  Ga and As atoms in  consecutive GaAs (001) atomic 
layers, the resulting configuration is self-compensating and corresponds to a 
{\em microscopic capacitor}
 with unique electronic properties.\cite{interlayers1,interlayers2}
A self-compensating layer configuration is required by electrostatic
stability, whereas its confinement and the value of the resulting dipole 
are limited by atomic inter-diffusion.
We study the simplest possible configurations, 
i.e. an entire  Si (001) bilayer confined onto two adjacent atomic planes
and the case of one monolayer uniformly distributed over two such planes.
We also study isolated surface donor-acceptor pairs 
which  can be considered the building blocks of 
extended self-compensating donor-acceptor configurations.
In all the cases XSTM images have been simulated using the
electronic-structure data resulting from our calculations.

The study of cross-sectional surfaces is conceptually similar 
to that of natural surfaces and can be afforded using  supercells with
slab geometries.\cite{note-slab}
Our numerical approach is based on the plane-wave pseudopotential
method  in the framework of density functional theory (DFT)
within the local-density approximation (LDA), using the ESPRESSO/PWscf
code.\cite{PWSCF}

The calculated structural properties of GaAs, bulk and along (110) surface,
are of the same good quality as currently found in the literature.
The calculated energy gap for bulk GaAs is 1.36 eV,
slightly underestimated with respect to the experimental value, 
as it typically occurs in DFT-LDA calculations.
The clean surface is semiconducting, with a calculated energy gap 
smaller than the bulk  value by about 0.5 eV.
The surface geometry is optimized by allowing all the atomic position
to relax.
STM images are obtained with the simple and widely used
model by Tersoff and Hamann.\cite{TersoffHamann} 
According to this model, the tunneling current is proportional to the
local density of states (LDOS) calculated
at the position of the tip and integrated in the energy range between 
the Fermi energy  $E_F$ and $E_F+eV_b$, where $V_b$ 
is the applied bias.
In this model the proximity of the tip is assumed not to perturb
the electronic structure of the surface.

The position of the calculated Fermi energy depends sensitively on the
concentration of dopants which is contrivedly large, for it
depends on the size of the supercell.
It is therefore appropriate to use the experimental value of the
Fermi energy in order to simulate
XSTM images from our calculated electronic-structure data.
The experimental samples are overall $n$-doped, and 
 the Fermi energy is thus close to the bottom 
of the  conduction band. Taking into account that our calculated surface
gap is approximately 0.3 eV smaller than the experimental value,
the experimental data at $V_b=-$1.7 V and $V_b=$1.3 V 
have to be compared with 
our calculations at  $V_b\approx -$1.5 V and $V_b\approx$~1.2 V
respectively.


We mainly study  neutral configurations.
However, in the Si-doped samples that we consider,
diffused as well as localized doping together with the presence of the surface
may induce some charge accumulation. This
may affect band structure and DOS not only 
by rigidly shifting them with respect to the Fermi energy 
but also by modifying their shape. For these reasons
we also examine some charged states of the samples and
we investigate their effects on the structural and electronic properties 
in a selfconsistent way. 

Our tests on  the clean (110) surface
and with the isolated $\rm Si_{\rm Ga}$ surface substitutional 
impurity confirm the findings of previous
experimental and theoretical 
investigations,\cite{Ebert-review,WangPRB93}
thus  validating our approach.
We have therefore considered  the possible scenarios
arising  from a progressively higher Si doping of GaAs
in the hypotesis of donor configurations ($\rm Si_{\rm Ga}$) only:
two surface $\rm Si_{\rm Ga}$ impurities with different relative positions
(on-surface or sub-surface) up to an entire 
row (again on-surface or sub-surface) or even an entire ML, 
abrupt or spread over two adjacent cationic planes.
The simulated XSTM images for all these c
onfigurations---not shown here---do not correspond to the bright features
at negative bias and attenuated contrast at positive bias  observed in
the recent experiments: they would
rather look the opposite.


We then consider a uniform distribution of substitutional $\rm Si_{\rm Ga}$
and  $\rm Si_{\rm As}$ atoms
over two adjacent cationic and anionic (001) layers.
The intersection of the (110) surface with the Si (001) bilayer can occur
in different positions and thus result in different configurations
of the exposed surface  that can include rows of Si atoms in As 
and/or Ga sites.
There are four different possible configurations:
(A) both $\rm Si_{\rm Ga}$ and  $\rm Si_{\rm As}$ on-surface, forming
a zig-zag chain; (B) both $\rm Si_{\rm Ga}$ and  $\rm Si_{\rm As}$
sub-surface, forming again 
a zig-zag chain; (C)  $\rm Si_{\rm Ga}$ atoms  on-surface and 
 $\rm Si_{\rm As}$ sub-surface, forming a row of  $\rm Si_{\rm
   Ga}$---$\rm Si_{\rm As}$ 
(001)--oriented pairs; (D) the complementary configuration with 
 $\rm Si_{\rm As}$ atoms  on the surface and  $\rm Si_{\rm Ga}$ in subsurface.
According to our calculations, configuration B is energetically slightly 
favoured with respect to the others. In Fig. 2 we report a ball and stick 
model of this structures, together with the two corresponding simulated
STM images.
We can see a very bright signal at negative
bias voltages which is strongly attenuated at positive voltages.
With lower resolution, the bright signal would appear as a line of 
1--1.5 nm wide, i.e. of dimension comparable with those reported in 
Ref.~\onlinecite{ModestiPRL04}.
The charge state has little influence 
on the XSTM image (results not shown here).
The other cases, A, C and D, are similar to this, apart from
the negative charge state of C and D, where
a weakly bright signal remains also at positive voltages.

Coming to 
a Si single monolayer uniformly distributed over two adjacent (001)
atomic layers---i.e. with one substitutional impurity every two atoms---the
overall features
found for the full bilayer also show in this case.
In particular, the bright/attenuated contrast at negative/positive
bias remains a characteristic feature also in configurations
with a reduced local concentration of Si dopants, which are more
likely to occur in real samples.

In order to better characterize the microscopic origin of this
feature, we have simulated the XSTM images of the point dipoles that
would result from isolated $\rm Si_{\rm Ga}$-$\rm Si_{\rm As}$
pairs on the exposed surface. In Fig. 3 we show  the case where
the donor and the acceptor lie in different on-surface and sub-surface
atomic layers. Note that the bright spots of $\approx$0.3 nm  wide 
at negative bias visible in this figure correspond to an
individual impurity confined to the exposed surface, whereas in the
previous cases (A--D) they were representative of an entire doping
layer shadowed by the surface spot.
The resulting image is
much less contrasted than that corresponding to the full bilayer,
 but the overall appearance is similar.
The different contrast between the images of the full
bilayer and those of the isolated dipole indicate therefore that also 
the Si atoms in the inner layers contribute to the 
modification of the local density of states and of the XSTM image. 

Summarizing,  we have shown that self-compensating
donor-acceptor layer configurations of Si dopants
are characterized by bright features at negative bias which are strongly 
attenuated by reversing it, in close resemblance with
experimental images of Si-doping layers embedded in GaAs samples.
We believe that in spite of the uncertainties due to 
several factors (tunneling voltage,
nature and shape of the tip in case of 
experiments, specific technical details  of the numerical simulations)
not fully under control, our findings provide a qualitatively correct picture
of the microscopic mechanism responsible for the observed images, thus
pointing to the value of first-principles simulations as a tool for the 
characterization of materials.

This work was supported, in part, by the
Italian National Institute for the Physics of Matter (INFM)
under the project PRA-XSTMS and 
the ``Iniziativa Trasversale di Calcolo
Parallelo''. 
We gratefully acknowledge  useful  discussions with
A. Franciosi, S. Rubini and M. Piccin from INFM-TASC National Laboratory in 
Trieste.
One of us (M.P.) also acknowledges  support for computational facilities
provided by the University of Trieste under 
  the agreement with the Consorzio Interuniversitario CINECA.



\newpage

\begin{figure}
  \includegraphics[angle=0,width=.45\textwidth]{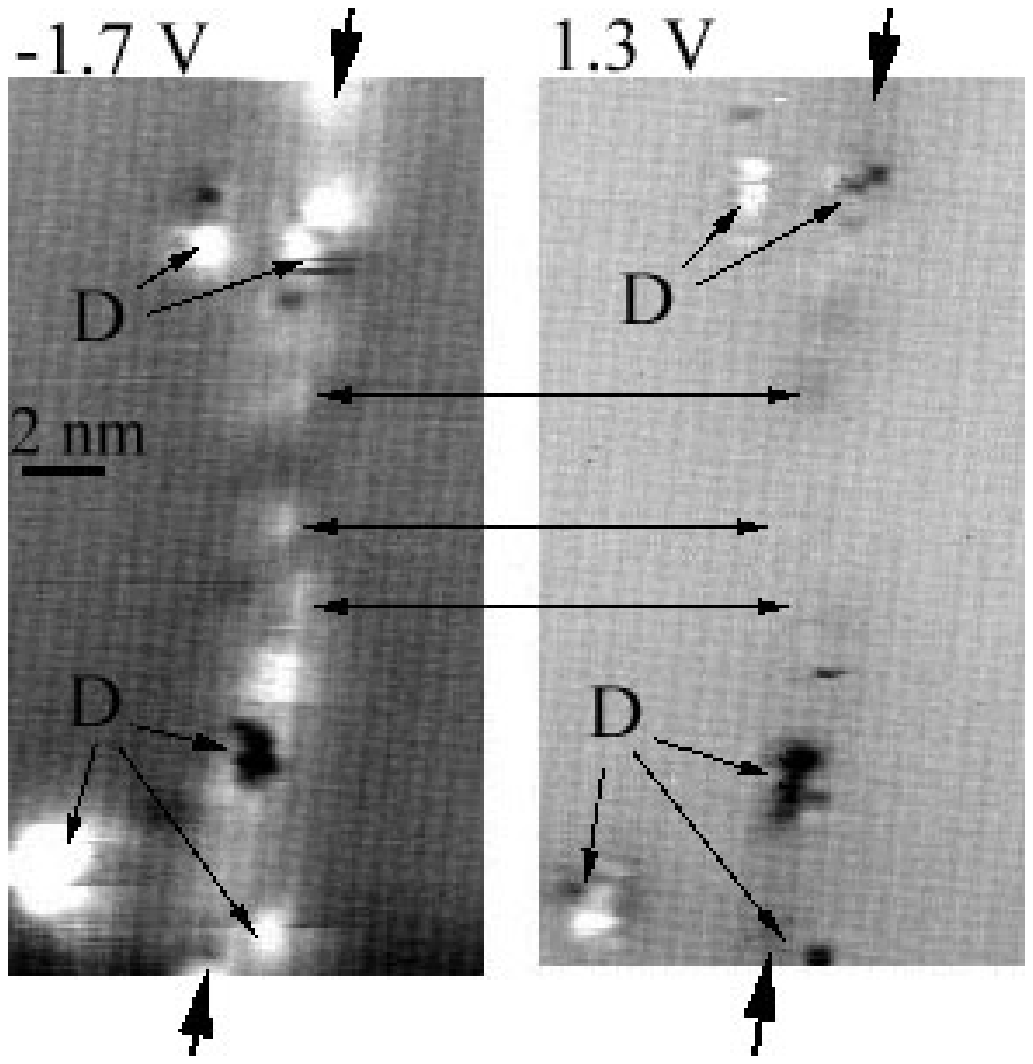}
\caption{
XSTM images of the same region of the intersection
between 2 ML of Si in GaAs parallel to the (001) plane and the (110)
cleavage surface taken at a sample bias of $-$1.7 V (left) and 1.3 V (right).
The tunneling current is 0.2 nA. The intersection line is marked by vertical
arrows. Orizontal arrows connect corresponding points of the Si layer.
D indicates cleavage defects. The Si
layer was grown by MBE at 600$^\circ$C and analysed by XSTM with the methods
described in Ref.~\protect{\onlinecite{ModestiPRL04}}.}  
\label{fig:exp}
\end{figure}
\clearpage
\newpage

\begin{center}
\begin{figure}
   \includegraphics[angle=0,width=.4\textwidth]{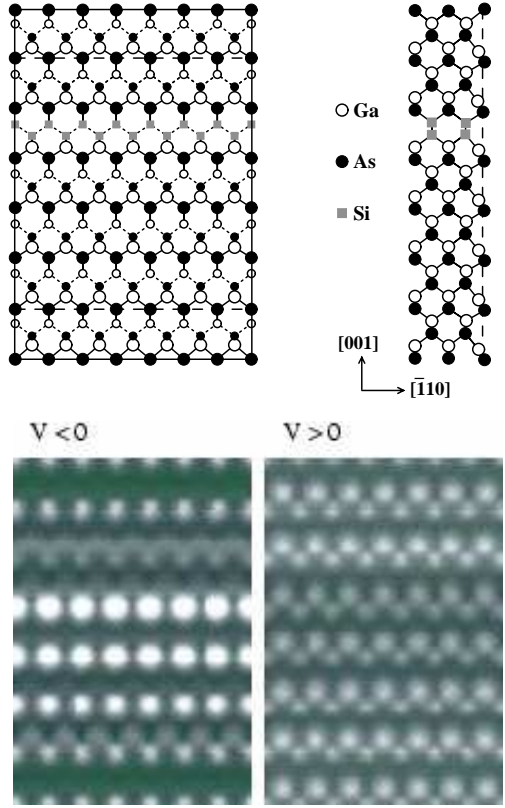}
\caption{(001) Si bilayer in GaAs: 
case with $\rm Si_{\rm Ga}$ and  $\rm Si_{\rm As}$ sub-surface
(case B in the text).
Top panels: ball and stick model of the relaxed surface,
top and side view (Ga: open circle, As: close circle, Si: square). 
The inner cell indicated with dashed lines in the 
top view is used to simulate the STM images 
of the defected region, which is superimposed to the image of the otherwise
perfect clean surface for a better visual rendering.
Bottom panels: simulated STM images at bias voltages of
$-$1.5 eV (occupied states) and at +1.2 eV (empty states) respectively.
The size of the entire region shown in our simulated STM images 
corresponds to $\approx$ 2.75 nm $\times$ 3.89 nm.
}
\label{fig:clu_B}
\end{figure}
\end{center}

\clearpage
\newpage

\begin{figure}
   \includegraphics[angle=0,width=.6\textwidth]{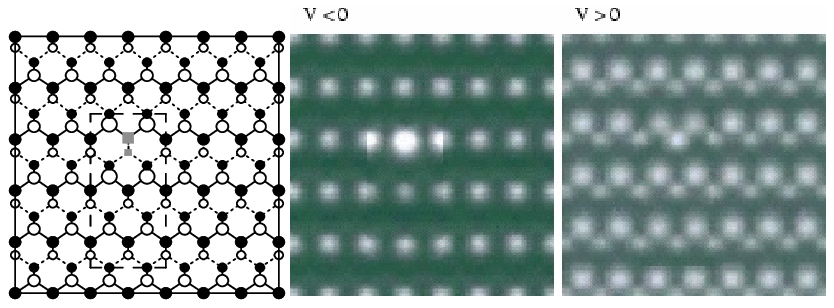}
\caption{Isolated $\rm Si_{\rm Ga}$--$\rm Si_{\rm As}$ 
pairs exposed on (110) GaAs surface:
$\rm Si_{\rm As}$ on surface, $\rm Si_{\rm Ga}$ subsurface.
See Fig. 2 for caption.}
\label{fig:dipoles}
\end{figure}


\begin{thebibliography}{00}
\bibitem{note-slab}
The (110) slab geometry is obtained by applying
periodic boundary conditions in the (1$\bar{1}$0), (110) and (001) directions.
The former is used to describe repeated parallel slabs (containing
9 atomic planes perpendicular to the (1$\bar{1}$0) direction in our case),
 separated by a vacuum region (of about 
12 {\AA} in our case) large enough  to guarantee
 a sufficient separation of two adjacent surfaces.
The last two periodic boundary conditions are used to describe
 the infinite surface; the surface unit cell  for the doped configurations
 has to be large enough (up to 20 atoms in the present cases)
in order to guarantee a sufficient separation 
between periodic images of the impurities.
\end{thebibliography}
\end{document}